\documentclass[aps,preprint,twocolomn,pacs]{revtex4}
\usepackage{graphicx}
\usepackage{dcolumn}
\usepackage{bm}

\begin{document}
\draft
\date{\today}
\title{S-matrix theory for transmission through billiards
in tight-binding approach}
\author{ Almas F. Sadreev$^{1,2}$ and Ingrid Rotter$^3$}
\affiliation{1)  Kirensky Institute of Physics, 660036,
Krasnoyarsk, Russia\\ 2) Department of Physics and Measurement
Technology, Link\"{o}ping University, S-581 83 Link\"{o}ping,
Sweden\\ 3) Max-Planck-Institute f\"ur Physik Komplexer Systeme,
D-01187 Dresden, Germany}
\begin{abstract}
In the tight-binding approximation we consider multi-channel
transmission through a billiard coupled to leads. Following Dittes
we derive the coupling matrix, the scattering matrix and the
effective Hamiltonian, but take into account the energy
restriction of the conductance band. The complex eigenvalues of
the effective Hamiltonian define the poles of the scattering
matrix. For some simple cases, we present exact values for the
poles. We  derive also the condition for the appearance of double
poles.
\end{abstract}
\maketitle
\section{Introduction}
In recent years, the ballistic transport through quantum systems
has been studied as a scattering problem on billiards (microwave
cavities) with infinitely high potential walls (hard wall
approximation). The scattering properties of such billiards are
closely related to the spectral properties of the corresponding
closed billiards \cite{doron,stockmann}. The opening of the
billiards is realized by attaching at least one lead to them.
However, for the study of the transmission through the billiard
two leads are necessary. The fundamental object that characterizes
the process of quantum scattering is the unitary S-matrix relating
the amplitudes of incoming waves to the amplitudes of outgoing
waves. Provided that the properties of the Hamiltonian $H_B$ for
the closed billiard are known, one can consider its open
counterpart and work out the S-matrix formalism by standard
methods of the theory of quantum scattering
\cite{feshbach,fano,weidenmuller,rotter,fyodorov,dittes,rotter1}.
As a result, the S-matrix is expressed in terms of both the
Hamiltonian $H_B$ and the matrix elements describing the coupling
of the billiard states to the lead states. The explicit
expressions for the coupling matrix elements were formulated at
first in 1996 by  \v{S}eba et al. \cite{seba,albeveiro} for the
case of point contacts of the leads with the billiard. Later,
Fyodorov and Sommers \cite{fyodorov} developed their theory for
the connection of the billiard  with one lead of finite width by
using Neumann boundary conditions (see also \cite{stockmann}).
Recently Dittes \cite{dittes} considered the same type of an open
system and derived the expressions for the coupling matrix with
Neumann or Dirichlet boundary conditions both by using the Green
function technique.

In the present paper, we consider a d-dimensional billiard
connected to leads by using another approach that is based on the
tight-binding model. The motivation for this consideration is the
following. First of all, the increasing development of fabrication
techniques requires the possibility to perform reliable numerous
experiments on ballistic transport through devices of atomic size
\cite{ohnishi,yanson} and through molecular devices consisting of
very few atoms. Secondly, Pichugin {\it et al} \cite{pichugin} who
applied the formula for the coupling matrix derived by Dittes
\cite{dittes}, found that their results do not coincide with those
of a direct numerical computation of the S-matrix poles, above all
for the Dirichlet boundary conditions. As we will show in the
present paper, one of the reasons for this disagreement is that
the formal continuum approach used by Dittes \cite{dittes} is
unbounded in energy and gives zero radiation shifts. In electron
transmission through electron wires, however, the energy of the
electrons is bounded in energy, at least from below. This fact
gives rise to radiation shifts of poles of the S-matrix which can
not be neglected in calculations for concrete systems. Thirdly, it
is desirable   to receive numerical results from an S-matrix
computation within the tight-binding model in order to compare
them with the results of numerical computation of the transmission
through billiards. To this aim we derive, in the present paper,
the coupling matrix, the effective Hamiltonian and the poles of
the S-matrix within the tight-binding model and present some
typical numerical results.

Here, the following remark should be added. The computer
simulations solve the Schr\"odinger equation using
finite-difference Hamiltonians, i.e. the tight-binding
approximation. After matching the incoming and outgoing waves with
the solutions of the Schr\"odinger equation by implying the
boundary conditions at the transverse sections of the leads, the
conductance of the billiard as well as the scattering wave
function can be computed. Today, the calculations can be performed
with a very high accuracy by using large grids and the technique
of sparse matrices. The current S-matrix theory is  adequate to
these computer simulations. It is, however, numerically  more time
consuming because the effective Hamiltonian is not a sparse
matrix. Nonetheless, calculations with the effective Hamiltonian
are useful since they provide another view to the results. In this
formalism, the resonant peaks of the conductance are related to
the poles of the S-matrix that correspond to the eigenvalues of
the effective Hamiltonian.  It is possible therefore to draw some
conclusions on the origin of the resonant peaks and on their
possible control by means of external parameters.

In the present paper we will follow, as closely as possible, the
Dittes review \cite{dittes}, even in the notations. In the case of
microwave or quantum semiconductor billiards, the waves are
incident to the billiard through  (infinitely) long straight
waveguides (leads) of a certain width. The different channels
correspond therefore to different transverse modes of the wave
propagation within the leads \cite{stockmann}. At a given
frequency $E$ (the Fermi energy), we enumerate the propagating
modes by $p=1,..., M$. Thereby, we associate with the lead region
a continuous set of states $|C, p, E>$ where $C$ specifies the
lead number (terminal). In the present paper, we consider mostly
two leads, the right lead with incident  and reflected waves and
the left lead with outgoing waves as shown in Fig. \ref{fig1}. Our
approach can, however, be easily generalized to a larger number of
leads as done in Sections 4 and 5.
\section{One-dimensional tight-binding model of resonant tunneling}
A numerical scheme for the computation of quantum transport
through billiards with attached leads is mainly based on the
finite-difference Schr\"odinger equation. Implying the Ando
procedure \cite{ando} for the boundary conditions has enabled us
to find the transmission properties of the billiards from the
scattering wave function for any geometry of billiards and
straight leads. In order to compare the results from such a
computation with the S-matrix theory, we consider systems
projected on a lattice with finite grid.

\begin{figure}[t]
\includegraphics[width=.6\textwidth]{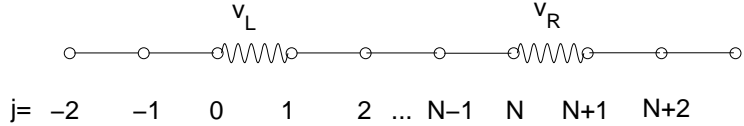}
\caption{The one-dimensional tight-binding model. The wave lines
couple the left and right leads with the box containing  $N$
points and, correspondingly,  $N$ resonant states.} \label{fig1}
\end{figure}

As a first example, we consider a simple one-dimensional model for
quantum scattering and transport. This model is formulated as the
tight-binding model (the Anderson model)
\begin{equation}\label{tight}
 H=-\sum t_j |j><j+1| + c.c.
\end{equation}
where $t_j=v_L$ if $j=0$, $t_j=v_R$ if $j=N$ , and $t_j=1$
otherwise. This tight-binding model presents the simplest case of
a one-dimensional box with $N$ sites coupled with left and right
semi infinite leads via the corresponding coupling constants $v_L,
v_R$. For a  schematic representation see Fig. \ref{fig1}.
 
\begin{figure}[t]
\includegraphics[width=.6\textwidth]{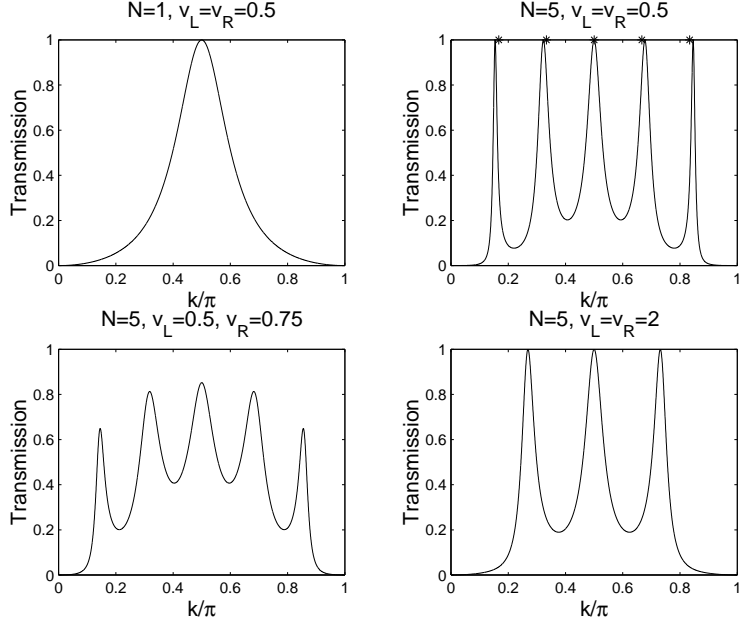}
\caption{The transmission probability versus the wave number of
the incident quantum particle. The positions of the eigenvalues of
the closed billiard ($v_L=~v_R=~0$) are shown by stars. One can
see the radiation shifts caused by the coupling of the 1d box with
the leads.}
 \label{fig2}
\end{figure}

The hopping matrix elements $t_j$ in the Hamiltonian (\ref{tight})
are proportional to overlapping integrals of electron wave functions
between adjacent atoms. 
The model being similar to the one-dimensional model with a
double barrier structure \cite{birula,azbel}, describes   resonant
tunneling.   
In this case, the values of the coupling coefficients
$v_L, v_R$ play the role of the heights of the double barrier
structure provided that  $v_L <1, v_R < 1$. The present model
gives, however,  also the possibility to consider the case of
strong coupling,  $v_L >1 , v_R > 1$. At the left of the box we
present the solution of the Schr\"odinger equation
\begin{equation}\label{schred}
H|\psi> = E|\psi>
\end{equation}
as
\begin{equation}\label{left}
\psi_j=e^{ikj}+re^{-ikj}, j<1
\end{equation}
where $r$ is the reflection coefficient with energy
\begin{equation}\label{energy}
E(k)=-2\cos k, -\pi \leq k \leq \pi.
\end{equation}
As  will be seen later, the energy $E(k)$ forms the  conduction
band $-2\leq E \leq 2$ with finite width. At the right of the box
we write
\begin{equation}\label{right}
\psi_j=te^{ikj}, j>N.
\end{equation}
At last, the solution of (\ref{schred}) inside the box  is
\begin{equation}\label{box}
\psi_j=ae^{ikj}+be^{-ikj}, j=1, 2,\ldots,N.
\end{equation}
Substituting these functions into (\ref{schred}) we obtain the
following linear equations
\begin{eqnarray}\label{system}
r(e^{ik}-E)+v_L e^{ik}a+v_L b e^{-ik}=e^{ik}\nonumber\\ v_L r+
e^{ik}(e^{ik}-E)a+ e^{-ik}(e^{-ik}-E)b=-v_L\nonumber\\
e^{ikN}(e^{-ik}-E)a+
e^{-ikN}(e^{ik}-E)b+v_Re^{ik(N+1)}t=0\nonumber\\
v_Re^{ikN}a+v_Re^{-ikN}b+e^{ik(N+1)}(e^{ik}-E)t=0.
\end{eqnarray}
The coefficients $a$ and $b$ can be expressed via the
transmission coefficient $t$ as follows
\begin{eqnarray}
a=t\frac{v_R -\frac{1}{v_R}e^{-2ik}}{1-e^{-2ik}},\nonumber\\
b=t\frac{(v_R -\frac{1}{v_R})e^{2ikN}}{1-e^{-2ik}} .
\end{eqnarray}
Finally, one obtains
\begin{eqnarray}\label{rt}
t=4sin^2k/A\\
r=\frac{t}{v_L(1-e^{-2ik})}\left[v_R-\frac{1}{v_R}e^{-2ik}+
(\frac{1}{v_R}-v_R)e^{2ikN}\right]-1\nonumber
\end{eqnarray}
for the solution of the system of equations (\ref{system}) where
$$ A=(v_L-\frac{1}{v_L})(v_R-\frac{1}{v_R})e^{2ikN}-e^{-2ik} (v_L
e^{2ik}-\frac{1}{v_L})(v_R e^{2ik}-\frac{1}{v_R}). $$ For the
particular case $v_L=v_R=1$, we obtain from (\ref{rt}) that $t=1$
and $r=0$. A typical resonant transmission through a
one-dimensional box is shown in Fig. \ref{fig2}.

The tight-binding model (\ref{tight}) demonstrates a few
remarkable features. The first one is the symmetry of the resonant
transmission relative to $v_{L,R}\rightarrow 1/v_{L,R}$. This
symmetry means the following: for small as well as for large
coupling coefficients, the effective  coupling of the box to the
leads is small. Such a feature was firstly observed in reactions
on atomic nuclei, see the review \cite{rotter}, and analytically
derived by Dittes {\it et al} for an $N$-level system coupled to
one open channel \cite{dittes1}. With increasing coupling
strength, the widths of $N-1$ resonance states decrease as $1/v$
in the single-channel case while only one resonance state
accumulates almost the total sum of the widths. In our case, the
$N$-level system is coupled to two open channels. Correspondingly,
with increasing coupling coefficients $v_L, ~v_R$ the widths of
$N-2$ resonance states decrease while the widths of two  resonance
states increase. We will return to this feature below when the
poles of the scattering matrix will be considered. The second
feature is that the heights of the resonant peaks are equal to one
only when $v_L=v_R$, similar to the double barrier resonant
structure. This fact was firstly established by Ricco and Azbel
\cite{azbel}. The radiation shifts of the positions of the
resonant peaks relative to the eigenvalues of the box
\begin{equation}\label{eigen1d}
E_n=-2cosk_n, ~k_n=\pi n/(N+1),~ n=1, 2, \ldots, N,
\end{equation}
are the third peculiarity of the tight-binding model. The
positions of the eigenvalues (\ref{eigen1d}) are shown in Fig.
\ref{fig2} by stars. The shifts and widths of the resonant peaks
are symmetrical relative to $k\rightarrow -k$, and $E\rightarrow
-E$.  The last symmetry follows from the invariance of the
solution of the tight-binding model relative to $t_j\rightarrow
-t_j$.

\section{The S-matrix for the 1d tight-binding model}
The simplicity of the model (\ref{tight}) allows us to establish
the explicit correspondence between the analytical results for the
transmission amplitudes (\ref{rt}) and the S-matrix approach
\cite{feshbach,fano,weidenmuller,rotter,dittes}. This 1d model was
also used in \cite{terraneo} to investigate the width
distribution. In this approach the scattering system is decomposed
into a closed subsystem described by the internal Hamiltonian
$H_B$ with discrete bound states $|\psi_n>, n=1, 2,...,N$ and the
continuum of external scattering states $|E,L>$ and $|E,R>$
corresponding to the semi infinite left and right leads. The
Hamiltonian of the two  uncoupled subsystems is
\begin{eqnarray}\label{H0}
H_0=H_B+H_L+H_R,\nonumber\\ H_B=\sum_n E_n |n><n|,\nonumber\\
H_L=\int_{-2}^2 dE E|E,L><E,L|, \quad H_R=\int_{-2}^2 dE
E|E,R><E,R|] ,
\end{eqnarray}
where $E_n$ are the energies of the bound states of the 1d closed
billiard, and $E$ denotes the energy of the leads. We use the following
normalization conditions
\begin{eqnarray}\label{norma}
<n|m>=\delta_{nm}\nonumber\\
<E,L|E',L>=<E,R|E',R>=\delta(E-E').
\end{eqnarray}

The couplings between the internal and external subsystems can be
incorporated by the coupling operator
\begin{equation}\label{coupling}
V=\sum_n \sum_{C=L,R}\int_{-2}^2 dE V_n(E,C) |E,C><n| +H.C.
\end{equation}

As shown in Fig. \ref{fig1} the closed 1d billiard consists of $N$
sites with energies given by Eq. (\ref{eigen1d}) and the corresponding
eigenfunctions
\begin{equation}
\label{psiB1d} \psi_n (j)=\sqrt{\frac{2}{N+1}}\sin\left(\frac{\pi
n j}{N+1}\right), ~j=1, 2, \ldots , N.
\end{equation}
These eigenfunctions satisfy  the Dirichlet boundary conditions
$\psi_n(0)=\psi_n(N+1)=0$. Since the leads are semi-infinite wires
the wave functions of the left and right lead are respectively
\begin{equation}\label{psileads}
\psi_{E,L}(j)=\sqrt{\frac{1}{2\pi |\sin k|}} \sin k(1-j),\quad
\psi_{E,R}(j)=\sqrt{\frac{1}{2\pi |\sin k|}} \sin k(j-N).
\end{equation}
The energy in the leads corresponding to a single conductance
energy band, is defined by $E(k)=-2\cos k,\quad -\pi\leq k \leq
\pi$. It is easy to see that in the continual limit $k\rightarrow
0$ the functions (\ref{psileads}) take the form
\begin{equation}\label{psileadcont}
\psi_{E,L}(x)=\sqrt{\frac{1}{2\pi|k|}} \sin kx
\end{equation}
given in \cite{fyodorov}. From (\ref{coupling}) we have $$
V_n(E,C)=<E,C|V|n>=<E,C|\sum_j|j><j|V|\sum_{j'}|j'><j'|n>,~C=L,R.
$$ Since as shown in Fig. \ref{fig1} the coupling matrix elements
$<j|V|j'>$ are not equaled to zero if only $j=0,1,~ j'=1,0$ or
$j=N,N+1, ~j'=N+1, N$, we obtain finally, using (\ref{psiB1d}) and
(\ref{psileads}), the coupling coefficients as
\begin{eqnarray}
\label{couplconst}
V_n(E,L)=v_L\psi_{E,L}(0)\psi_n(1)=v_L\sqrt{\frac{\sin k}{\pi
(N+1)}}\sin{\frac{\pi n}{N+1}},\nonumber\\
V_n(E,R)=v_R\psi_n(N)\psi_{E,R}(N+1)=v_R\sqrt{\frac{\sin k}{\pi
(N+1)}} \sin\frac{\pi nN}{N+1}.
\end{eqnarray}

For the total Hamiltonian
\begin{equation}\label{H}
H=H_0+V
\end{equation}
the stationary Schr\"odinger equation reads
\begin{equation}\label{shred1}
H|\psi(E)>=E(k)|\psi(E)>.
\end{equation}
For $E(k)$ different from the eigenvalues $E_n$ of the box, the operator
$(E+i0-H_0)V$ is well defined and  equation (\ref{shred1}) is
equivalent to the Lippmann-Schwinger equation
\begin{equation}\label{LS}
|\psi>=|\psi_0>+(E+i0-H_0)^{-1}V|\psi>
\end{equation}
if the boundary condition of outgoing waves  is adopted and
\begin{equation}\label{k}
(E-H_0)|\psi_0>=0.
\end{equation}
The Lippmann-Schwinger equation (\ref{LS}) also reads
\begin{equation}\label{LS1}
|\psi>=[F(E+i0)]^{-1}|\psi_0>,
\end{equation}
where
\begin{equation}\label{F}
F(E+i0)=1-(E+i0-H_0)^{-1}V
\end{equation}
and
\begin{equation}\label{psi0}
|\psi_0>=\left(\begin{array}{c} |E,L> \cr
                            0 \cr
       |E,R> \end{array}\right).
\end{equation}

Following to \cite{weidenmuller,dittes} we introduce three
projection operators: for the left and right leads
\begin{equation}\label{PLR}
P_C=\int dE |E,C><E,C|
\end{equation}
and for the billiard
\begin{equation}\label{PB}
P_B=\sum_n |n><n|
\end{equation}
with the help of which we can write the scattering wave function (\ref{LS1})
as
\begin{equation}\label{psiproj}
|\psi>=\left(\begin{array}{c} P_L|\psi> \cr
    P_B|\psi> \cr
     P_R|\psi> \end{array}\right) =
     \left(\begin{array}{c} |\psi_L> \cr
    |\psi_B> \cr
     |\psi_R> \end{array}\right).
\end{equation}
Then the coupling operator (\ref{coupling}) reads
\begin{equation}\label{Vproj}
V=\left(\begin{array}{ccc} P_LVP_L & P_L V P_B & P_L V P_R \cr
P_BVP_L & P_B V P_B & P_B V P_R \cr P_RVP_L & P_R V P_B & P_R V
P_R \end{array}\right) =\left(\begin{array}{ccc} 0 & V_{LB} & 0
\cr V_{BL} & 0 & V_{BR} \cr 0 & V_{RB} & 0 \end{array}\right)
\end{equation}
where by using (\ref{couplconst})
 we obtain
\begin{eqnarray}
\label{VBL} V_{BL}=v_L\sum_n \psi_n(1)\sqrt{\frac{1}{2\pi}} \int
dE~ [1-(E/2)^2]^{1/4}|n><E,L|,\nonumber\\ V_{BR}=v_R\sum_n
\psi_n(N)\sqrt{\frac{1}{2\pi}} \int dE~ [1-(E/2)^2]^{1/4}|n><E,R|,
\end{eqnarray}
and $V_{BC}=V_{CB}^{+}$. Substituting (\ref{Vproj}) into (\ref{F})
we have
\begin{equation}\label{FF}
F=\left(\begin{array}{ccc} 1    & -\frac{1}{E-H_L}V_{LB} & 0 \cr
         -\frac{1}{E-H_B}V_{BL} & 1 & -\frac{1}{E-H_B}V_{BR} \cr
            0  &    -\frac{1}{E-H_R}V_{RB} &        1  \end{array}\right).
\end{equation}
Using the identity
\begin{equation}\label{identity}
\left(\begin{array}{ccc} 1    & -A & 0 \cr
          -B & 1 & -C \cr
             0  &    -D &         1  \end{array}\right)^{-1}=
\left(\begin{array}{ccc} 1+ATB    & AT & ATC \cr
        TB & T & TC \cr
DTB  & DT &   1+DTC   \end{array}\right)
\end{equation}
one obtains for the inverse matrix $F$
\begin{equation}\label{Finverse}
F^{-1}=\left(\begin{array}{lll}
1+\frac{1}{E-H_L}V_{LB}\frac{1}{D}\frac{1}{E-H_B}V_{BL} &
\quad\frac{1}{E-H_L}V_{LB}\frac{1}{D}  &
\quad\frac{1}{E-H_L}V_{LB}\frac{1}{D}\frac{1}{E-H_B}V_{BR} \cr
\frac{1}{D}\frac{1}{E-H_B}V_{BL} & \quad\frac{1}{D} &
\quad\frac{1}{D}\frac{1}{E-H_B}V_{BR} \cr
\frac{1}{E-H_R}V_{RB}\frac{1}{D}\frac{1}{E-H_B}V_{BL}   &
\quad\frac{1}{E-H_R}V_{RB}\frac{1}{D} & \quad
1+\frac{1}{E-H_R}V_{RB}\frac{1}{D}\frac{1}{E-H_B}V_{BR}
\end{array}\right).
\end{equation}
where $$ T=\frac{1}{1-BA-CD}$$ and
\begin{equation}
\label{D} D=1-\frac{1}{E-H_B}\sum_{C=L,R}
V_{BC}\frac{1}{E-H_C}V_{CB} .
\end{equation}
From Eqs (\ref{LS1}) and (\ref{psiproj}) it follows that the
 wave function in the interior of the billiard
 is
\begin{equation}
\label{psibilliard} |\psi_B>=Q^{-1}\sum_{C=L,R} V_{BC}|E,C>
\end{equation}
where
\begin{equation}
\label{Q} Q=E^{+}-H_B-\sum_{C=L,R}
V_{BC}\frac{1}{E^{+}-H_C}V_{CB}.
\end{equation}
Here we used the indentity \cite{dittes} $$
\frac{1}{1-AB}A=A\frac{1}{1-BA}. $$

If we substitute the coupling constants (\ref{couplconst}) into
formula (\ref{Q}), it follows that the matrix elements of the
operator (\ref{Q}) can be presented as matrix elements of the
effective Hamiltonian \cite{fano,dittes}
\begin{equation}
\label{Q1}
<m|Q|n>=E^{+}\delta_{mn} - <m|H_{eff}|n>
\end{equation}
where
\begin{equation}
\label{Heff}
<m|H_{eff}|n>=E_m\delta_{mn}+\frac{1}{2\pi}V_{mn}\int_{-2}^2 dE_1
\frac{\sqrt{1-(E_1/2)^2}}{E+i0-E_1} =E_m\delta_{mn}-V_{mn} e^{ik}.
\end{equation}
The last expression was obtained by using the formula $$
\frac{1}{x+i0}=i\pi\delta(x)+{\it P}\frac{1}{x}$$ where ${\it P}$
denotes the principal value integral and
\begin{equation}
\label{Vmn}
V_{mn}=v_L^2\psi_m(1)\psi_n(1)+v_R^2\psi_m(N)\psi_n(N).
\end{equation}
As can be seen from (\ref{Heff}), the lattice approach gives rise to a
finite shift of the resonant energies
\begin{equation}
\label{shift}
F_{mn}(E)=\frac{1}{2N}V_{mn}E
\end{equation}
which is of the same order of magnitude as the  width of the resonant peak
of the transmission
\begin{equation}
\label{width}
\gamma_{mn}(E)=\frac{1}{N}V_{mn}\sqrt{1-(E/2)^2}.
\end{equation}
This energy shift is the main difference between the present
tight-binding (lattice) approach and the continuum approach by
Dittes \cite{dittes} where the shifts are equal to zero. The
reason for this difference is that  the energy is restricted to
the conductance band $E=-2\cos k$.

The S-matrix is \cite{weidenmuller,dittes}
\begin{equation}
\label{Smatrix}
S_{CC'}=\delta_{CC' }-2\pi
i<E,C|V_{CB}Q^{-1}V_{BC'}|E,C'>=\left(\begin{array}{cc} r & t'\cr
t & r' \end{array}\right)
\end{equation}
where $r$ and $r'$ are the reflection coefficients from left to
left and from right to right, respectively, and $t$ and $t'$ are
the transmission coefficients from left to right and from right to
left. Using the definition of the coupling operator (\ref{VBL}) we
can write the transmission coefficient of the S-matrix
(\ref{Smatrix}) as follows
\begin{equation}
\label{transmission} t=-2\pi
i\sum_{mn}V_m(E,L)<m|Q^{-1}|n>V_n^{*}(E,R).
\end{equation}
A concrete calculation of the transmission coefficient needs the
procedure of inversion of the matrix (\ref{Q1}). It is therefore
more convenient to use a representation by means of the set of
eigenstates of the effective Hamiltonian (\ref{Heff})
\cite{rotter,rotter1}. Using the biorthogonal basis of the
effective Hamiltonian
\begin{equation}
\label{eigHeff}
 H_{eff}|\lambda)=z_{\lambda}|\lambda),~
(\lambda|\lambda')= \delta_{\lambda,\lambda'},\quad
|\lambda)=|\lambda>, ~(\lambda| = <\lambda|^{*},
\end{equation}
and the projection operator
\begin{equation}
\label{Peff} P_{eff}=\sum_{\lambda}|\lambda)(\lambda|
\end{equation}
we obtain
\begin{equation}
\label{transmissionheff} t=-2\pi
i\sum_{\lambda}\frac{<E,L|V|\lambda)(\lambda|V|E,R>}{E-z_{\lambda}}.
\end{equation}
Eq. (\ref{transmissionheff}) shows immediately that the
eigenvalues of the effective Hamiltonian ~$z_{\lambda}$~ define
the poles of the scattering matrix. We underline that the coupling
coefficients in the pole representation (\ref{transmissionheff})
have to be calculated by means of the eigenstates $|\lambda)$ of
the effective Hamiltonian but not with the eigenstates $|b>$ of
the closed billiard. The importance of this difference is
presented in \cite{persson}.

Let us consider for illustration the limiting case $N=1$, the
one-sided Anderson model \cite{terraneo}, which corresponds to the
1d box with a single eigenstate. For simplicity we take
$v_L=v_R=v$. Then the formula for the transmission coefficient
(\ref{rt}) reduces to
\begin{equation}
\label{trN1} t=-\frac{iv^2 \sin{k}}{\cos{k}-v^2 e^{ik}}.
\end{equation}
On the other hand, from (\ref{Heff}) and (\ref{Vmn}) we have the
effective Hamiltonian as the c-number $H_{eff}=z_1=E_1-2v^2e^{ik}$
where for the one-sided dot $E_1=0$. Moreover
$V_1(E,L)=\tilde{V}_1(E,R)=v\sqrt{\frac{\sin k}{2\pi}}$.
Substituting these formulas into (\ref{transmissionheff}) we
obtain the same formula as (\ref{trN1}) with account that
$E=-2\cos k$. An analysis of the S-matrix for the transmission
through the $N$-sided 1d box is given in Section 5.


\section{S-matrix theory for the transmission through billiards}

\begin{figure}[t]
\includegraphics[width=.6\textwidth]{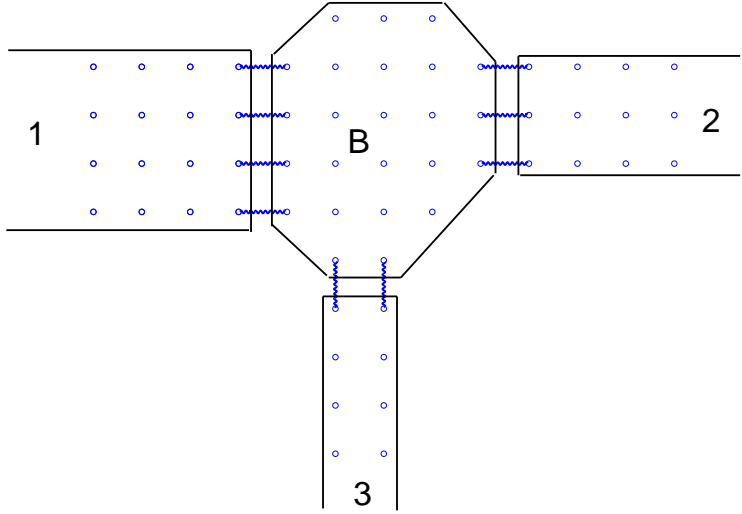}
\caption{The two-dimensional billiard $B$ attached to three different
leads $C=1,2,3$. The coupling coefficients $v_C$ between the leads and
the billiard are shown by wave lines}. \label{fig3}
\end{figure}

Let us consider a d-dimensional billiard  specified by the  internal
eigenstates $|b>$ and eigenvalues $E_b$,
\begin{equation}
\label{d-billiard}
H_B|b> =E_b|b>.
\end{equation}
The shape of the billiard is given by the $d-1$ surface $\Omega$
which encloses the internal region of the billiard $D$ with the
points ${\bf  x}\in D$. The eigenfunctions are $<{\bf
x}|b>=\psi_b({\bf x})$. We assume that $M$ leads are attached to
the billiard. Each lead is a d-dimensional tube with arbitrary
transverse section $\omega_C, ~C=1, 2,\ldots, M$ and is
semi-infinite along the direction $z\perp \omega_C$. The geometry
of the system is illustrated in Fig. \ref{fig3} for the
two-dimensional case and $M=3$. It allows a separation of
variables ${\bf x}_{\perp}\in \omega_C$ and $z$.

Assuming that the eigenvalues and eigenfunctions  for the
transverse section  of the C-th lead are known and denoted by
$E_{p_C}, ~\phi_{p_C}({\bf x}_{\perp})$, we can write the
Schr\"{o}dinger equation for the leads in the following manner
\begin{equation}
\label{dlead}
H_C|E,C,p_C> =E|E,C,p_C>.
\end{equation}
Here
\begin{equation}
\label{Edlead}
E=-2\cos(k_{p_C})+E_{p_C},
\end{equation}
\begin{equation}
\label{dleadpsi} \psi_{p_C}({\bf x})= \sqrt{\frac{1}{2\pi |\sin
k_{p_C}|}} \sin k_{_C}(j_z-j_C)\phi_{p_C}({\bf x}_{\perp}),
\end{equation}
and $j_C$ is the longitudinal position of the attachment of the
C-th lead to the billiard. Then the Hamiltonian of the uncoupled
system consisting of the billiard and the $M$ leads is
\begin{equation}
\label{H0d}
H_0=\sum_b E_b
|b><b|+\sum_{C=1}^M\sum_{p_C}\int_{-2+E_{p_C}}^{2+E_{p_C}} dE
E[|E,C,p_C><E,C,p_C|+H.C.
\end{equation}
Similar to (\ref{coupling}) let us write the coupling operator as
\begin{equation}
\label{couplingd}
 V=\sum_b \sum_C\sum_{p_C} \int_{-2+E_{p_C}}
^{2+E_{p_C}} dE V_b(E,C,p_C) |E,C,p_C><b| +H.C.,
\end{equation}
where
\begin{equation}
\label{Vd}
V_b(E,C,p_C)= <E,C,p_C|V|b>.
\end{equation}
Let be $A_C \subset \Omega$  the areas at which the leads are
attached to the billiard. They terminate the semi-infinite leads
at $j_z=j_C$. The shape of $A_C$ is  $\omega_C$,  the transverse
section of the lead. The C-th lead is connected to the billiard
through the hopping matrix elements $v_C$, as shown in Fig.
\ref{fig3} by wave lines. Substituting (\ref{dleadpsi}) into
(\ref{Vd}), we obtain for the coupling matrix elements
\begin{equation}
\label{meVd}
V_b(E,C,p_C)= \sum_{\bf x,y}\psi_{p_C}({\bf x})<{\bf
x}|V|{\bf y}>\psi_b({\bf y}) = v_C\sqrt{\frac{|\sin
k_{p_C}|}{2\pi}} \sum_{{\bf x}_{\perp}\in A_C}\phi_p({\bf
y}_{\perp})\psi_b({\bf y}_{\perp}).
\end{equation}

It is justified to generalize the  one-dimensional case presented
in Section 2 to the general case with $d>1$. In the following, we
present some formulas that follow in a straightforward manner.
Formulas (\ref{Q1}) and (\ref{Heff}) read now
\begin{equation}
\label{Qd}
 <b|Q|b'>=E^{+}\delta_{bb'}-<b|H_{eff}|b'>
\end{equation}
where
\begin{equation}
\label{Heffd}
 <b|H_{eff}|b'>=E_b\delta_{bb'}-\sum_C
\sum_{p_C} W_C(b,p_C)W_C(b',p_C)e^{ik_{p_C}}
\end{equation}
and
\begin{equation}
\label{Wd} W_C(b,p)=v_C\sum_{{\bf x}_{\perp}\in A_C}\psi_b({\bf
x}_{\perp})\phi_p({\bf x}_{\perp}).
\end{equation}
The number of channels in each lead is defined by the condition
$E_{p_C} < E$. In the continual case the energy is much less than
the width of the energy propagation band equaled to ~4. Therefore
we can approximate $e^{ik}\approx -E/2+i$. As a result the
effective Hamiltonian (\ref{Heffd}) takes the standard form
\begin{equation}
\label{Heffcont}
 H_{eff}=\tilde{H}_B-iWW^{+},
\end{equation}
where $\tilde{H}_B$ is the billiard Hamiltonian the eigenenergies
of which are corrected by the radiation shifts, and $W$ is the
matrix whose elements are given by (\ref{Wd}). The dimension of
the matrix $W$ is $N\times K$ where $N$ is the number of states in
the billiard, $K=\sum_C \max(p_C)$.

Finally, the S-matrix elements (\ref{Smatrix}) are characterized
by the channel numbers and read
\begin{equation}
\label{Smatrixd} <C,p_C|S|C',p'_{C'}>=\delta_{CC'}\delta_{pp'}-
2\pi i<E,C,p_C|V_{CB}Q^{-1}V_{BC'}|E,C',p'_{C'}>.
\end{equation}
For calculation of the S-matrix we can use the set of the
eigenstates $\psi_b$ of the billiard (\ref{transmission}) or the
biorthogonal set of eigenstates $\psi_{\lambda}$ of the effective
Hamiltonian (\ref{transmissionheff}). In the last case one can see
immediately that the poles of the S-matrix correspond to the
eigenvalues of the effective Hamiltonian (\ref{Heffd}).


\section{Some applications of the general theory}

\subsection{Transmission through a 2d rectangular billiard}

\begin{figure}[t]
\includegraphics[width=.6\textwidth]{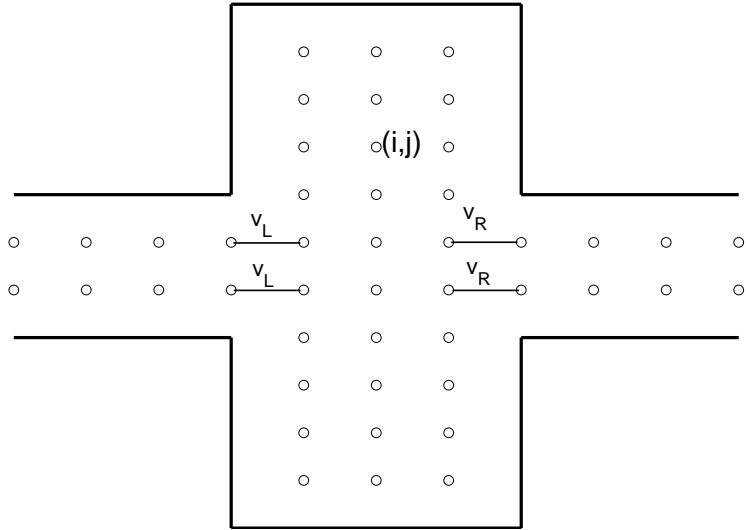}
\caption{The geometry for the transmission through a rectangular
billiard in the tight-binding approach. The couplings $v_C$
between the leads and the billiard are shown by solid lines}
\label{fig4}
\end{figure}

The typical features of the quantum mechanical transmission
through a billiard can be described by means of a two-dimensional
billiard with two attached leads of equal widths (Fig. \ref{fig4}).
In the tight-binding formulation  ${\bf x}=a_0(i,j)$
where $a_0$ is the lattice unit. The eigenfunctions and eigenvalues
of the rectangular billiard are
\begin{eqnarray}
\label{psiB2d}
\psi_{m,n}(i,j)=\psi_m(i)\psi_n(j)\\
\label{Emn}
E_{m,n}=E_m+E_n
\end{eqnarray}
where $E_m$ and $\psi_m$ are defined by equations (\ref{eigen1d}),
(\ref{psiB1d}) for the corresponding numerical size of the box
$N_x, N_y$. Further, the wave functions (\ref{dleadpsi}) of the
leads and their energy (\ref{Edlead}) are
\begin{eqnarray}
\label{psilead2d} \psi_{E,L,p}(i,j)=\sqrt{\frac{1}{2\pi |\sin
k_p|}} \sin k_p(1-i) \phi_p(j),\nonumber\\
\psi_{E,R,p}(i,j)=\sqrt{\frac{1}{2\pi |\sin k_p|}} \sin k_p(i-N_x)
\phi_p(j) ,
\end{eqnarray}
\begin{equation}
\label{energylead2d}
E=-2\cos k_p +E_p,~ E_p = -
2\cos\left(\frac{\pi p}{N_L+1}\right)
\end{equation}
where $p=1, 2, 3, ...$ enumerates the channel number and $N_L$ is
the numerical width of the leads. We denote the numerical
positions of the lead's walls  by $N_1$ and $N_2$ so that
$N_L=N_2-N-1$. It follows from the geometry of the system  that
$1\leq N_1 \leq N_2 \leq N_y$. Therefore, the area of intersection
between leads and  billiard $A_C$ is  a straight line of length
$N_L$. The eigenfunctions in the transverse sections of the leads
have the following form
\begin{equation}
\label{philead2d}
\phi_p(j)=
\sqrt{\frac{2}{N_L+1}}\sin\left(\frac{\pi p(j-N_1)}{N_L+1}\right).
\end{equation}
Substituting (\ref{psiB2d}) and (\ref{philead2d}) into
(\ref{meVd}) we have for the elements of the coupling matrix
\begin{eqnarray}
\label{meV2d}
V_{m,n}(E,L,p)= v_L\psi_m(1)\sqrt{\frac{|\sin
k_p|}{2\pi}} \sum_{j=N_1}^{N_2}\phi_p(j)\psi_n(j),\nonumber\\
V_{m,n}(E,R,p)= v_R\psi_m(N_x)\sqrt{\frac{|\sin k_p|}{2\pi}}
\sum_{j=N_1}^{N_2}\phi_p(j)\psi_n(j).
\end{eqnarray}
Here the Latin indexes $L, R$ denote the left and right leads,
respectively, as shown in Fig. \ref{fig4}. The formulas for the
effective Hamiltonian (\ref{Heffd}) with (\ref{Wd}) read
\begin{eqnarray}
\label{Heff2d}
<m,n|H_{eff}|m^{'},n^{'}>=E_{m,n}\delta_{mm^{'}}\delta_{nn^{'}}
-v_L^2\psi_m(1)\psi_{m'}(1)\sum_p^{\Lambda}
W_L(n,p)W_L(n^{'},p)e^{ik_p} \nonumber\\
-v_R^2\psi_m(N_x)\psi_{m'}(N_x)\sum_p^{\Lambda}
W_R(n,p)W_R(n^{'},p)e^{ik_p},
\end{eqnarray}
with
\begin{equation}
\label{W2d} W_C(n,p)=\sum_{j=N_1}^{N_2}\psi_n(j)\phi_p(j).
\end{equation}
The number of channels $\Lambda$ is defined by the condition
$\epsilon_p < E$.

Let us now consider the correspondence of the formulas obtained
to those received in the continuum approach by Dittes \cite{dittes}.
First  it is necessary  to choose, in the last case,  some
characteristic space length. This may be the width of the lead
or the size of the billiard. Here we choose, as usually in the literature,
the former and denote it by $d$. In the
continuum approach, the eigenfunctions and the eigenvalues of the
rectangular billiard (\ref{psiB2d}) take the following form
\begin{eqnarray}
\label{psiBcon}
\tilde{\psi}_{m,n}(x,y)=\tilde{\psi}_m(x)\tilde{\psi}_n(y),\quad
\tilde{\psi}_m(x)=\sqrt{\frac{2}{a}}\sin(mx/a)\\
\label{tildeEmn}
\tilde{E}_{m,n}=E_0\pi^2\left\{\frac{m^2}{(a/d)^2}+\frac{n^2}{(b/d)^2}\right\}
\end{eqnarray}
for Diriclet boundary conditions,
where $a$ and $b$ characterize the size of the billiard and
$E_0=\hbar^2/2md^2$. The eigenfunctions and eigenenergies of the
leads are
\begin{equation}
\label{psileadcon}
\psi_{E,L,p}(x,y)=\sqrt{\frac{1}{\pi d |k|}} \sin kx \sin \pi py/d
\end{equation}
\begin{equation}
\label{tildeenergylead}
\tilde{E}=E_0\left[(\tilde{k}d)^2+(\pi
p)^2\right] \, .
\end{equation}
With $x=a_0 i,~y=a_0 j$ we  find the following relations
\begin{equation}
\label{correspondence}
1+N_x=a/a_0,~ 1+N_y=b/a_0,~1+N_L=d/a_0
\end{equation}
from the comparison of
(\ref{psiB2d}) and (\ref{psilead2d}) with (\ref{psiBcon}) and
(\ref{psileadcon}).
In the discrete case $\psi_m(j=0)=0$. Therefore,
it holds approximately
$$\psi_m(1)=a_0\frac{\psi_m(1)-\psi_m(0)}{a_0}=a_0\psi_m'(0)$$
for the continuum case, and
the coupling matrix elements (\ref{Vd}) are
\begin{equation}
\label{V2dcont}
V_{m,n}(E,L,p)=V_0\Psi_{m,n,p}^{'}(0)
\end{equation}
where $$ \Psi_{\alpha,p}(x,y)=\int_{y_1}^{y_2} dy\phi_p(y)
\psi_{m,n}(x,y), ~~~V_0= \sqrt{\frac{1}{2\pi k_p}}\, ,$$ and $y_1, ~y_2$
are the positions of the lead walls along the y-axis, so that $d=y_2-y_1$.
The same expressions were derived in \cite{dittes,seba}.

It might  seem that, for the continuum limit $a_0 \rightarrow 0$,
the radiation shifts go to zero and we can use the
Weidenm\"uller-Dittes  approach directly. However as can be seen
from (\ref{tildeenergylead}),  the energy  is bounded from below.
As a consequence, the principal value integral in the matrix
elements of the effective Hamiltonian does not vanish. This is in
difference to the assumption in \cite{dittes} that  $\tilde{E}$
has  no limits.

\subsection{Transmission through a two-sided quantum dot}

A box consisting of a single atom (site) coupled to a left lead
and a right one (L and R continuous) is the most simple case that
gives rise to the Breit-Wigner type formula for the transmission
amplitude (\ref{trN1})  shown in Fig. \ref{fig2}~(a). Let us now
consider a two sites box that  gives rise to a $2\times 2$
effective Hamiltonian. The properties of such  Hamiltonians are
studied in literature \cite{rotter1} by focusing onto double poles
of the S-matrix (branch points in the complex energy plane)
without relation to a realistic system as well as in relation to
laser induced structures in atoms \cite{magunov}. In our present
study, we have the possibility to specify the effective $2\times
2$ Hamiltonian for another specific system  to study its
properties and to compare the results with those of the general
study.

\begin{figure}
\includegraphics[width=.4\textwidth,height=0.1\textheight]{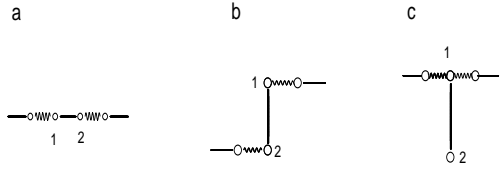}
\caption{Two-sided dot coupled to two leads.} \label{fig5}
\end{figure}

There are different ways to connect  the two cite box with the leads
as shown in Fig. \ref{fig5}. The cases (a) and (b) are identical,
but differ from (c).

\begin{figure}
\includegraphics[width=.6\textwidth,height=0.5\textheight]{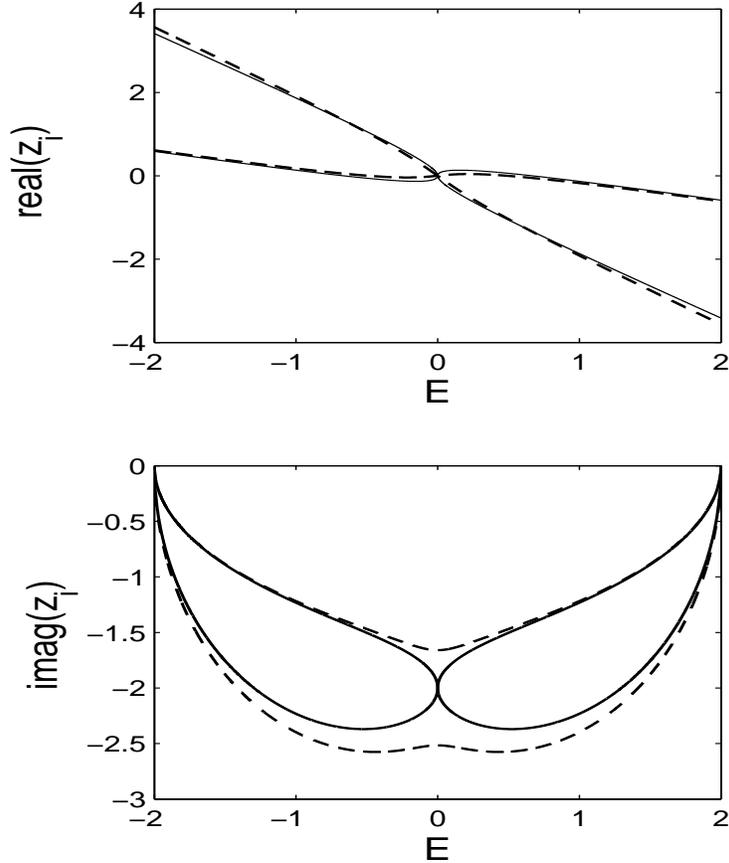}
\caption{The energy behavior of the poles for the transmission
through the two-sided dot shown in Fig. \ref{fig5}(a). (a) and
(b): $|\mu|=1+0.03$. (c) and (d): $|\mu|=1-0.03$ (dashed curves).
The solid curves in (c) and (d) show the case of a double pole,
$|\mu|=1.$} \label{fig6}
\end{figure}

For the case (a) we obtain from (\ref{Heff2d})
\begin{equation}
\label{Heffab} H_{eff} =-\left(\begin{array}{cc} -1+\lambda &\mu
\cr \mu & -1+\lambda \cr
\end{array}\right),
\end{equation}
where
\begin{equation}
\label{lammu}
 \lambda=\frac{1}{2}(v_L^2+v_R^2)e^{ik},\quad \mu= \frac{1}{2}
 (v_L^2-v_R^2)e^{ik}.
\end{equation}
The eigenvalues are:
\begin{equation}
\label{eigab} z_{1,2}=-\lambda \pm \sqrt{1+\mu^2}.
\end{equation}
They define the poles of the S-matrix as shown in Sections 3 and
4. Since the effective Hamiltonian is symmetric but not Hermitian
we use the biorthogonal basis \cite{moiseyev,rotter} normalized by
the condition $(m|n)=\delta_{mn}, ~m = 1,2, ~n = 1,2$ where
$(m|\equiv <m|^{*}$. The  right eigenstates of (\ref{Heffab}) are:
\begin{equation}
\label{eigfunab} |1>=\left(\begin{array}{c} a_1 \cr
     a_2 \cr
    \end{array}\right) = \frac{1}{\sqrt{2\eta(\eta+1)}}\left(\begin{array}{c} -\mu \cr
     1+\eta \cr
    \end{array}\right),\quad
     |2>=\left(\begin{array}{c} a_2 \cr
      -a_1 \cr \end{array}\right).
\end{equation}
With the formulas (\ref{Heffd}), (\ref{Wd}) the
transmission amplitude takes the  form
\begin{equation}
\label{trN2} t=\frac{2iv_Lv_R \sqrt{1-(E/2)^2}}{(E-z_1)(E-z_2)}.
\end{equation}
The S-matrix has a double pole when two of the eigenvalues (\ref{eigab})
coincide, i.e. when
\begin{equation}
\label{doubleab}
 E=0, \quad |\mu|=1.
\end{equation}
The energy behavior of the poles (\ref{eigab}) is shown in Fig.
\ref{fig6}. Such a kind of pole behaviour was shown in many works
based on the general presentation of the effective Hamiltonian as
a $2\times 2$ matrix (see, for example, review \cite{rotter1}).
The cases (a) and (b) in fig. \ref{fig6} ($|\mu| >1$) correspond
to a free crossing of energy levels in the complex plane, while
the cases (c) and (d) (dashed curves) correspond to the
self-avoided crossing.

The case (c) in Fig. \ref{fig5} gives
\begin{equation}
\label{Heffc}
H_{eff} =\left(\begin{array}{cc} -1+\lambda &\lambda
\cr    \lambda & 1+\lambda \cr
    \end{array}\right),
\end{equation}
where $\lambda$ is given by (\ref{lammu}). The poles of the
scattering matrix are
\begin{equation}
\label{eigc}
z_{1,2}=\lambda \pm \sqrt{1+\lambda^2}
\end{equation}
and the transmission amplitude is given by
\begin{equation}
\label{trN2c}
t=-\frac{2iv_Lv_R
E\sqrt{1-(E/2)^2}}{(E-z_1)(E-z_2)}.
\end{equation}
A double pole of the S-matrix can be found at
\begin{equation}
\label{doublec} E=0, \quad |\lambda|=1.
\end{equation}

The comparison of (\ref{trN2c}) with (\ref{trN2}) shows that the
way the leads are connected  with the box, plays an important role
for the conductance. In particular, the connection (a) gives rise
to a transmission zero only at the edges of the energy band $E =
\pm 2$ while  the connection (c) leads to $t = 0$ at $E = 0$. The
energy behavior of the poles (\ref{eigc}) is, however, similar to
that of the poles (\ref{eigab}) and a double pole of the S-matrix
appears at $E=0$ in both cases. Therefore, the transmission is
equal to zero in case (c) at the energy where the S-matrix has a
double pole while this is not so in the cases (a) and (b). This
shows clearly that the transmission zero for the case (c) is an
interference effect.

\subsection{Transmission through the N-sided 1d box}


As a next application, we consider the 1d model with $N$ sites
presented in Fig. \ref{fig1} in order to understand the reduction
of the number of transmission peaks by enlarging the coupling
coefficients. In Fig. \ref{fig2}, the resonant transmission has
$N=5$ peaks at small coupling coefficients $v_L, v_R$ but $N-2$
peaks for large coupling coefficients.
\begin{figure}
\includegraphics[width=0.75\textwidth,height=0.5\textheight]{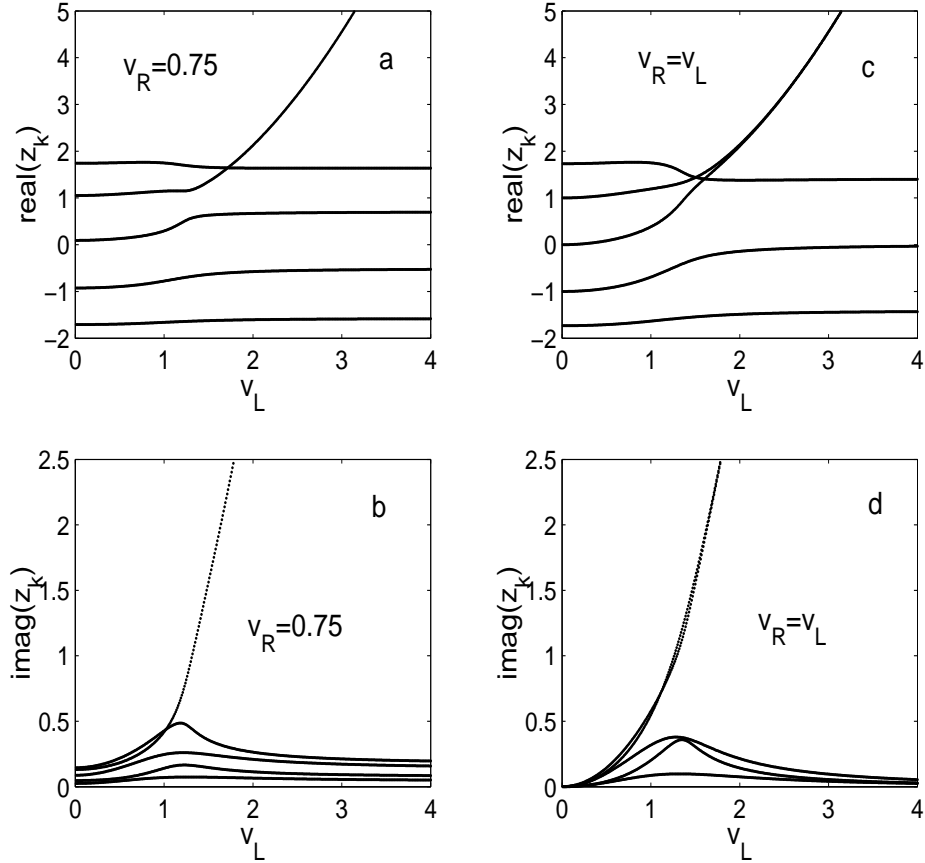}
\caption{Real and imaginary parts of the five poles of the 1d
chain shown in Fig. \ref{fig1} versus the coupling coefficients
$v_L, ~v_R,~ E = 1$. The chain consists of five sites.}
\label{fig7}
\end{figure}

Using formulas (\ref{Heff}) and (\ref{Vmn}), the  eigenvalues
$z_k, k=1, 2, \ldots, N$ of the effective Hamiltonian can be found
numerically. In Fig. \ref{fig7}, the real and imaginary parts of
the five eigenvalues of the effective Hamiltonian (poles of the
S-matrix) are shown versus the coupling constants $v_L, ~v_R$ for
$N = 5$ and $E = 1$. In Fig. \ref{fig7}(a, b), the right coupling
coefficient $v_R$ is chosen to be small. In this case, one of the
resonance states is broadened with increasing $v_L$ and becomes
shifted  beyond the energy band. The incident energy $E$ is tuned
to the second energy level $E_2=1$ of the box. As a result, this
resonance state is broadened. Fig. \ref{fig7} (c, d) demonstrates
that two resonance states  are broadened when both coupling
constants are increased. Also in this case, the two resonance
states are shifted beyond the energy band. Such a nonuniform level
broadening in the resonance overlapping regime is studied in many
different cases by using different approaches, see \cite{rotter1}.
It is called resonance trapping \cite{rotter}. The accompanying
shift in energy appears only when the principal value integral of
the matrix elements of the effective Hamiltonian is non-vanishing.
This is the case in most realistic systems including atomic nuclei
and atoms \cite{rotter1} and also those considered in the present
paper (Fig. \ref{fig7} (a, c)).

Thus, the two resonance states do not vanish at strong coupling
strength between box and leads as it might be concluded from Fig.
\ref{fig2}. The two resonance states go beyond the energy band and
can therefore contribute to the transmission only via interference
with the remaining narrow resonant states. This example clearly
demonstrates the advantage of the effective Hamiltonian approach
to the description of transmission.

\subsection{Transmission through a 2d billiard connected to  1d leads}

\begin{figure}
\includegraphics[width=.75\textwidth,height=0.3\textheight]{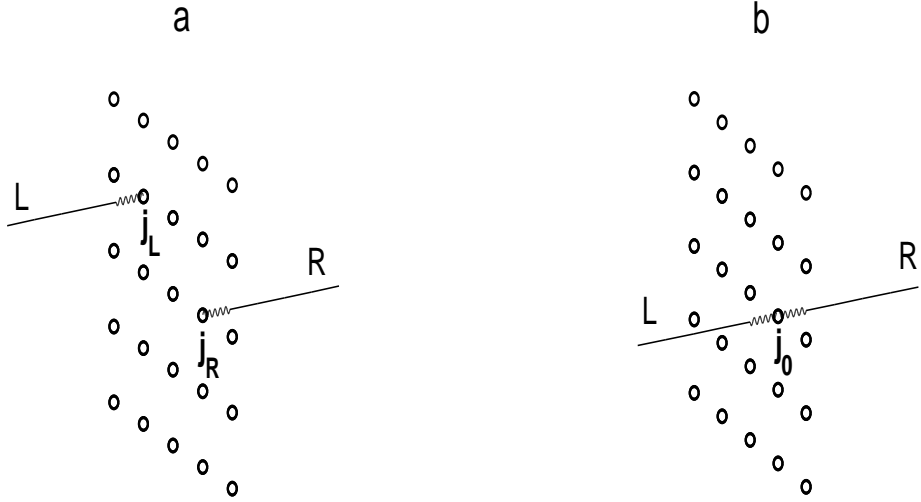}
\caption{The billiard coupled to  left and right 1d leads at the
points ${\bf j}_L$ and ${\bf j}_R$ (a). In (b)  ${\bf
j}_L={\bf j}_R = {\bf j}_0$. For simplicity a rectangular
billiard is shown in the figure (although it may be of arbitrary shape). The
couplings between the billiard and leads $v_L$ and $v_R$ are shown
by wave lines.}
\label{fig8}
\end{figure}

Let two 1d leads be  coupled to a 2d billiard at the points ${\bf
j}_L$ (input lead) where ${\bf j}_R$ (output lead) and  ${\bf j}=
(j_x, j_y)$,  $N_z=1$ as shown in Fig. \ref{fig8} (a).
Substituting the wave functions of the 1d
leads (\ref{psileads}) we obtain from Eq. (\ref{Heffd})
\begin{equation}
\label{Heff2d1} <b|H_{eff}|b'> = E_b\delta_{bb'} +
[v_L^2\psi_b({\bf j}_L) \psi_{b'}({\bf j}_L) +v_L^2\psi_b({\bf
j}_R) \psi_{b'}({\bf j}_R)]e^{ik}.
\end{equation}
For ${\bf j}_L={\bf j}_R$, the equation that defines the poles of
the scattering matrix, can be found analytically. From
(\ref{Heff2d1}) it follows
\begin{equation}
\label{determinant} \left|\begin{array}{llll}
E_1+\omega\psi_1^2({\bf j}_0) -E & \quad \omega\psi_1({\bf j}_0)
\psi_2({\bf j}_0) & \quad \omega\psi_1({\bf j}_0) \psi_3({\bf
j}_0) & \ldots \cr \omega\psi_1({\bf j}_0) \psi_2({\bf j}_0) &
\quad E_2+\omega\psi_2^2({\bf j}_0) -E & \quad \omega\psi_1({\bf
j}_0) \psi_3({\bf j}_0) & \ldots \cr \omega\psi_1({\bf j}_0)
\psi_3({\bf j}_0) & \quad \omega\psi_2({\bf j}_0) \psi_2({\bf
j}_0) & \quad E_3+\omega\psi_3^2({\bf j}_0) -E & \ldots \cr \vdots
& \quad \vdots & \quad \vdots & \ldots \cr
\end{array}\right| = 0,
\end{equation}
where $\omega=(v_L^2+v_R^2]e^{ik}$ is the effective coupling
constant. The particular case of a $4\times 4$  effective
Hamiltonian (\ref{determinant}) was considered in \cite{rotter2}.
This determinant can easily be transformed to \cite{sadreev}
\begin{equation}
\label{determinant1}
\prod_b \omega \psi_b^2({\bf j}_0)\left|\begin{array}{cccc} x_1+1
& 1 & 1 & \ldots \cr 1 & x_2+1 & 1 & \ldots \cr 1 & 1 & x_3+1
 & \ldots \cr \vdots & \vdots & \vdots
& \ldots \cr\end{array}\right| = \prod_b \omega \psi_b^2({\bf
j}_0)\left\{1+\sum_b \frac{1}{x_b}\right\} = 0,
\end{equation}
where $$x_b=\frac{E_b-E}{\omega \psi_b^2({\bf j}_0)}.$$  As a
result, the equation for the poles of the S-matrix reads
\begin{equation}
\label{npoles}
\sum_b \frac{(v_L^2+v_R^2)e^{-ik}\psi_b^2({\bf j}_0)}{E-E_b}=0.
\end{equation}

\subsection{A 3d billiard connected to a 3d lead}

Consider a 3d billiard that has an arbitrary shape in the $x, y$
plane but is restricted  in the z-direction  by two parallel
planes separated by a distance $d$. In the tight-binding
approximation, the height of the 3d billiard can be specified by
the number $N_z$ being equal to $1, 2, 3, \ldots $. This billiard
allows to separate  the variables, and it is characterized  by the
eigenvalues and eigenstates of the 3d box Hamiltonian $H_B$
\begin{equation}
\label{3dbilliard} H_B|b_{\perp},n_z>
=(E_{b_{\perp}}+E_{n_z})|b_{\perp},n_z>,
\end{equation}
where $E_{b_{\perp}}$ are the transverse eigenenergies of the
billiard and
\begin{equation}
\label{Enz} E_{n_z}=  -2\cos\left(\frac{\pi n_z}{N_z+1}\right)
    \quad n_z = 1, 2, \ldots, N_z
\end{equation}
are the longitudinal eigenenergies of the box in $z$-direction.
The eigenfunctions in the $z$-direction are
\begin{equation}
\label{psiz} \psi_{n_z}(z) =
    \sqrt{\frac{2}{N_z+1}}\sin\left(\frac{\pi n_z z_j}{N_z+1}\right), \quad
    j_z = 1, 2, \ldots , N_z.
\end{equation}

\begin{figure}
\includegraphics[width=.75\textwidth,height=0.3\textheight]{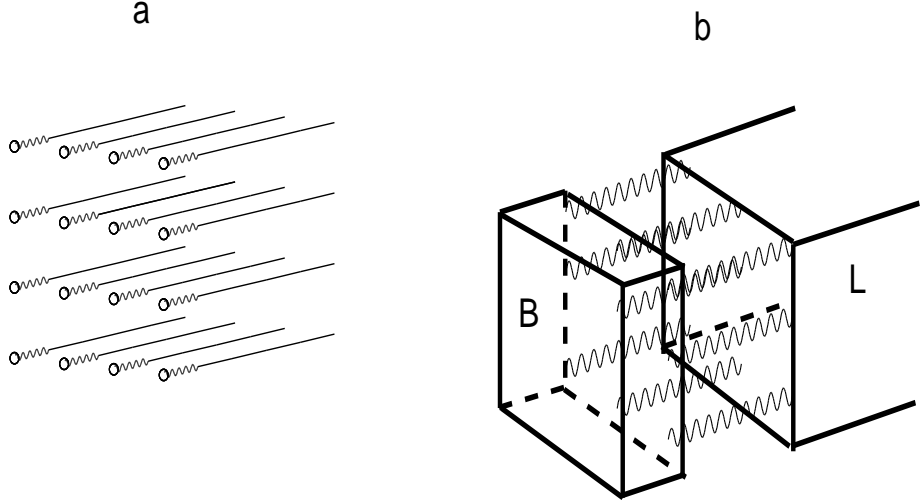}
\caption{3d billiards connected to different 3d leads. (a) The
billiard is coupled to $N_b$ 1d leads at each point ${\bf j}$
where $N_b$ is the number of sites of the billiard. (b) The
billiard is coupled to one 3d lead with the same transverse
section as the   billiard in 2d. For simplicity, a rectangular
billiard is shown although it can be of arbitrary shape. The
couplings $v$ between billiard and leads  are shown by wave
lines.} \label{fig9}
\end{figure}

We consider two different types of the connection between the
billiard and the leads that both are shown in Fig. \ref{fig9}. In
the first  case, (a), the 1d leads are coupled to the billiard at
every point ${\bf j}, j_z=1$ of the billiard. In the second case,
(b), the billiard is coupled to the 3d lead the transverse section
of which coincides with the shape of the billiard in 2d.

For the case (a) formula (\ref{Q}) reads
\begin{equation}
\label{QNb}
Q=E^{+}-H_B-\sum_{C=1}^{N_b}V_{BC}\frac{1}{E^{+}-H_C}V_{CB},
\end{equation}
where $N_b$ is the total number of cites of the billiard in the x,
y plane. This number  is equal to the total number of states
$|b>$. The coupling matrix elements are
\begin{equation}
\label{VNb} <b,n_z|V_{BL}|C>=\sum_{{\bf j}_1}\sum_{{\bf
j}_2}<b|{\bf j}_1><{\bf j}_1|V_{BC}|{\bf j}_2><{\bf
j}_2|C>=v\psi_b({\bf j}_C)\sqrt{\frac{|\sin k|}{2\pi}},
\end{equation}
where we assume that the coupling between the billiard and each 1d
lead (with the eigenfunction (\ref{psileads})) is equaled to $v$.
Substituting (\ref{VNb}) into (\ref{QNb}) we obtain, similar to
(\ref{Heff}),
\begin{equation}
\label{HeffNb} <b|H_{eff}|b'>=E_b\delta_{bb'}
+\sum_{C=1}^{N_b}\int dE_1
<b|V_{BC}|C>\frac{1}{E+i0-E_1}<C|V_{CB}|b'>=
(E_b+v^2e^{ik})\delta_{bb'}.
\end{equation}
That means,  for the case (a) in Fig. \ref{fig9} the effective
Hamiltonian is  diagonal with isolated poles $z_b=E_b+v^2e^{ik},
~k=a\cos(-E/2)$.

In the case (b) of Fig. \ref{fig9},  the billiard is coupled
to one 3d lead, and  the total Hamiltonian is
\begin{eqnarray}
\label{H3d}
H=H_0+V,\nonumber\\ H_0= \sum_{b,n_z} (E_b+E_{n_z})
|b,n_z><b,n_z|+\sum_b\int_ {-2+E_b}^{2+E_b} \int dE
E[|E,b><E,b|+|b,R><E,b|],\nonumber\\ V=\sum_{bb'}\sum_{nn_z'}
\int_{-2+E_{b'}} ^{2+E_{b'}} dE V_{b,n_z,b'}(E) |E,b'><b,n_z|
+H.C.,
\end{eqnarray}
where the eigenfunctions of the lead are
\begin{equation}
\label{psi3dlead}
<{\bf j},j_z|E,b>=\sqrt{\frac{1}{2\pi |\sin k_b|}} \sin k_b(1-j_z)
\psi_b({\bf j}) \, .
\end{equation}
Here, $j_z$ runs along the lead and  ${\bf j}$ runs over the sites
in the transverse section of the lead. Since the transverse
section eigenfunctions of the 3d lead coincide with those of the
2d billiard, similar to (\ref{energylead2d}), we have
\begin{equation}
\label{energylead3d} E=-2\cos k_b +E_b.
\end{equation}
The coupling matrix elements  in  (\ref{H3d}) are
\begin{equation}
\label{V3d}
  V_{b,n_z,b'}(E)=<b,n_z|V|E,b'>=\sum_{{\bf j},j_z} \sum_{{\bf l},l_z}
  <b|{\bf j}><n_z|j_z><{\bf j},j_z|V|{\bf l},l_z><{\bf l},l_z|E,b'>.
\end{equation}
As can be seen from Fig. \ref{fig9} $$<b|{\bf j},j_z><{\bf j}|V|{\bf
l},l_z>=v\delta_{{\bf j},{\bf l}}\delta_{j_z,N_z}\delta_{l_z,1}.$$
Substituting (\ref{psi3dlead}) into  (\ref{V3d}), we get
\begin{equation}
\label{V3dd}
  V_{b,n_z, b'}(E)=v\sqrt{\frac{|\sin k_{b'}|}{2\pi}} \sum_{{\bf j}}
  <b|{\bf j}><{\bf j}|b'><N_z|n_z> =v\sqrt{\frac{|\sin
  k_{b'}|}{2\pi}}\delta_{bb'}\psi_{n_z}(N_z).
\end{equation}
For the continual case the last expression has to be substituted by
$\psi'(z=d)$ \cite{dittes}.
Therefore the matrix elements of the effective Hamiltonian are
\begin{eqnarray}
\label{Heff3d}
  <b|H_{eff}|b'>=\left\{(E_b + E_{n_z})\delta_{n_zn_z'}+
  v^2\psi_{n_z}(N_z)\psi_{n_z'}(N_z)\frac{1}{2\pi}
  \int_{-2+E_b}^{2+E_b}dE'\frac{\sin
 k_b}{E+i0-E'}\right\}\delta_{bb'}\nonumber\\
 =[(E_b+E_{n_z})\delta_{n_zn_z'}
 +v^2e^{ik_b}\psi_{n_z}(N_z)\psi_{n_z'}(N_z)]\delta_{bb'}.
\end{eqnarray}
If $N_z=1$,  we obtain an effective Hamiltonian for  the case (b)
in Fig. \ref{fig9}, that is diagonal in the eigen basis of the
billiard. This result is similar to that of the case (a). For
$N_z>1$, the effective Hamiltonian is also diagonal, however with
blocks $N_z\times N_z$ at each diagonal place.  For $N_z=2$, with
account of (\ref{Enz}) and (\ref{psiz}), the matrix block takes
the following form
\begin{equation}
\label{Heff3db}
H_{eff} =\left(\begin{array}{cc}
E_b+1+\frac{v^2}{2}e^{ik_b} &\frac{v^2}{2}e^{ik_b} \cr
\frac{v^2}{2}e^{ik_b} & E_b-1+\frac{v^2}{2}e^{ik_b}\cr
    \end{array}\right).
\end{equation}
Correspondingly, the poles of the S-matrix are
\begin{equation}
\label{poles3db} z_{n_z,b}=E_b+\frac{v^2}{2}e^{ik_b}\pm \sqrt{1
+\frac{v^4}{4}e^{2ik_b}}.
\end{equation}
Using  (\ref{energylead3d}), the condition for the double pole
can be written down,
\begin{equation}
\label{doublepole3db}
 v^2=2,\quad E=E_b \, .
\end{equation}


\section{Wave function in the interior of the billiard}


The wave function in the interior of the billiard is given by the
expression (\ref{psibilliard}). Using the projection operator for
the billiard, $P_B=\sum_b |b><b|$ where $|b>$ are the eigenstates
of the (closed) billiard, we can rewrite this expression as
follows
\begin{equation}
\label{psiB}
\psi_B({\bf x})=\sum_{C,p_C}
\sum_{bb'}Q^{-1}_{bb'}V_{b'}(E,C,p_C)\psi_b({\bf x})=\sum_b f_b
\psi_b({\bf x}).
\end{equation}
The scattering wave function in the interior of the billiard can therefore be
expanded in the set of eigenfunctions $\psi_b({\bf x})$  of the
Hamiltonian of the closed billiard.  The expansion  coefficients are
\begin{equation}
\label{fb} f_b=\sum_{C,p_C} \sum_{b'}Q^{-1}_{bb'}V_{b'}(E,C,p_C).
\end{equation}

The drawback of this representation consists in the fact that the
expansion (\ref{psiB}) includes the procedure of inversion of the
matrix (\ref{Qd}). Similar to (\ref{transmissionheff}) we can use
the set of eigenfunctions of the effective Hamiltonian for the
expansion of the scattering wave function. Using relations
(\ref{eigHeff}) and (\ref{Peff}) we can write (\ref{psibilliard})
as follows
\begin{equation}
\label{psiBeff} \psi_B({\bf
x})=\sum_{\lambda}f_{\lambda}\psi_{\lambda}({\bf x}),
\end{equation}
where the  expansion coefficients are
\begin{equation}
\label{feff}
f_{\lambda}=\sum_{C,p_C}\frac{V_{\lambda}(E,C,p_C)}{E^{+}-z_{\lambda}}.
\end{equation}

\section {Summary}

In this paper we derived the coupling matrix between a closed
billiard and leads attached. The knowledge of the coupling matrix
gives the explicit expression for the effective Hamiltonian, the
S-matrix and the scattering wave function in the interior of the
billiard . The non hermitian effective Hamiltonian reflects the
spectral properties of the closed billiard. The eigenvalues of the
effective Hamiltonian however are shifted in energy and are
complex because of the openness of the billiard.

The theory presented is based on the tight-binding approach. That
allows us to establish the exact correspondence between the
S-matrix theory and numerical calculation of the transmission
through the billiard that is based on a finite-difference
Hamiltonian. The present approach can be easily applied to the
continual case. The advantage of the effective Hamiltonian
consists above all in the possibility to interpret numerical
results for the transmission (Fig. \ref{fig2}) by means of the
poles of the S-matrix (Fig. \ref{fig7}). The last are the
eigenvalues of the effective Hamiltonian. It allows us therefore
to systematically control the transmission through billiards. We
presented a few specific examples for which the effective
Hamiltonian reduces to a complex two by two matrix.

\acknowledgments We thank Konstantin Pichugin for discussions.
This work has been partially by RFBR Grant 01-02-16077,
03-02-17039 and the Royal Swedish Academy of Sciences. A.F.S
thanks also Max-Planck-Institute f\"ur Physik Komplexer Systeme
for hospitality.\\ $^{*}$ e-mails:almsa$@$ifm.liu.se,
almas$@$tnp.krasn.ru, rotter$@$mpipks-dresden.mpg.de

\end{document}